\documentstyle [11pt] {article}
\input epsf

\begin{document}
\oddsidemargin .175in
\begin{flushright}
\end{flushright}
\vspace {.5in}
\begin{center}
{\Large\bf Random matrix theory for mixed regular-chaotic\\ dynamics
in the super-extensive regime\\} \vspace{.5in}

{\bf A. Abd El-Hady ${}^{a,b,} \footnote{email : $alaa\_abdelhady@yahoo.com$
}$, A. Y. Abul-Magd ${}^{c,d,}\footnote{email : $ayamagd@hotmail.com$
}$\\}

\vspace{.1in} ${}^{a)}$ {\it
Department of Physics, King Khalid University, Abha, Saudi Arabia}\\
\vspace{.1in} ${}^{b)}$ {\it
Department of Physics, Zagazig University, Zagazig, Egypt}\\

\vspace{.1in} ${}^{c)}$ {\it
  Department of Mathematics, Zagazig University, Zagazig, Egypt}\\
 \vspace{.1in} ${}^{d)}$ {\it
Faculty of Engineering Sciences, Sinai University, El-Arish, Egypt}\\

\vskip .5in
\end{center}

\begin{abstract}

We apply Tsallis's $q$-indexed nonextensive entropy to formulate a
random matrix theory ({\bf RMT}), which may be suitable for
systems with mixed regular-chaotic dynamics. We consider the
super-extensive regime of $q<1$. We obtain analytical expressions
for the level-spacing distributions, which are strictly valid for
$2\times 2$ random-matrix ensembles, as usually done in the
standard {\bf RMT}. We compare the results with spacing
distributions, numerically calculated for random matrix ensembles
describing a harmonic oscillator perturbed by Gaussian orthogonal
and unitary ensembles.


\end{abstract}

\newpage


%
\section{INTRODUCTION\label{intro}}
%

The past decade has witnessed a growing interest in Tsallis'
non-extensive generalization of statistical mechanics
\cite{Ts1,Ts2}. The formalism has been successfully applied to a
wide class of phenomena; for a review see, e.g. \cite{Ts2,Ts23}. The
standard statistical mechanics is based on the Shannon
entropy measure $S=-\int dxf(x)\ln f(x)$ (we use Boltzmann's constant $%
k_{B}=1$), where $f(x)$ is a probability density function. This
entropy is extensive. For a composite system $A+B$, constituted of
two independent subsystems $A$ and $B$ such that the probability
$p(A+B)=p(A)p(B),$ the entropy of the total system
$S(A+B)=S(A)+S(B)$. Tsallis proposed a
non-extensive generalization: $S_{q}=\left. \left( 1-\int dx\left[ f(x)%
\right] ^{q}\right) \right/ (q-1)$. The standard statistical
mechanics is recovered when the entropic index $q$ = 1. The value
of $q$ characterizes the degree of extensivity of the system. The
entropy of the composite system $A+B$, the Tsallis' measure
verifies
\begin{equation}
S_{q}(A+B)=S_{q}(A)+S_{q}(B)+(1-q)S_{q}(A)S_{q}(B),
\end{equation}%
from which the denunciation non-extensive comes. Therefore, $%
S_{q}(A+B)<S_{q}(A)+S_{q}(B)$ if $q>1$. This case is called
sub-extensive. If $q<1$, the system is in the super-extensive
regime. The relation between the parameter $q$ and the underlying
microscopic dynamics is not fully understood yet.

Non-extensive generalization to the random matrix theory ({\bf
RMT}) \cite{mehta} has recently received considerable attention \cite{abul51,abul52}.
The conventional {\bf RMT} is a statistical theory of random
matrices $\rm{\bf{H}}$ whose entries fluctuate as independent
Gaussian random variables. The matrix-element distribution is
obtained by extremizing Shannon's entropy subject to the
constraint of normalization and existence of the expectation value
of Tr$\rm{\bf{(H^+H)}} $ \cite{balian}. Non-extensive
generalizations of {\bf RMT}, on the other hand, extremize
Tsallis' non-extensive entropy, rather than Shannon's. The first
attempt in this direction is probably due to Evans and Michael
\cite{evans}. Toscano et al. \cite{toscano} constructed
non-Gaussian ensemble by minimizing Tsallis' entropy and obtained
expressions for the level densities and spacing distributions. A
slightly different application
of non-extensive statistical mechanics to {\bf RMT} is due to Nobre and Souza \cite%
{nobre}. Recently, Bertuola et al. \cite{bertuola} generated a new
family of
ensembles that unifies some important applications of {\bf RMT}. In the range -$%
\infty <q<1$, this family was found to be a restricted trace
ensemble that interpolates between the bounded trace ensemble
(Chapter 19 of \cite{mehta}) at $q=-\infty $ and the Wigner
Gaussian ensemble at $q=1$. The non-extensive formalism was
applied in Ref. \cite{abul} to ensembles of $2\times 2$ matrices.
Analytical expressions for the level-spacing distributions of mixed
systems belonging to the orthogonal-symmetry universality class
are obtained in \cite{abul}. The calculation of the spacing
distribution showed different behavior depending on whether $q$ is
above or below 1. It is found that the sub-extensive regime of
$q>1$ \cite{abul1} correspond to the evolution of a mixed system
from chaos modelled by the standard {\bf RMT} to order described
by the Poisson statistics. On the other hand, the spectrum behaves
in a different way in the super-extensive regime, where $q\,<1$.
Starting with a value of $q=1$, where the nearest-neighbor spacing
({\bf NNS}) distribution is well approximated by the Wigner
surmise (see below), and decreasing $q$ the distribution becomes
narrower, and more sharply peaked at spacing $s=1$. It develops
towards the picked-fence type, such as the one obtained by Berry
and Tabor \cite{berry} for the two-dimensional harmonic oscillator
with non-commensurate frequencies.

The present paper considers the departure from chaos in the
super-extensive regime.  Section 2 reviews the non-extensive
formulation of {\bf RMT}. It contains the derivation of the level
spacing distributions for super-extensive systems with and without
time reversible symmetry. Section 3 is devoted to a numerical
experiment in which a two-dimensional harmonic oscillator is
perturbed by a random-matrix ensemble. It is shown that the
super-extensive {\bf RMT} provides a reasonable description of the
final stage of the stochastic transition. The conclusion of this
work is given in Section 4.

\section{Non-extensive {\bf RMT}}

{\bf RMT} replaces the Hamiltonian of the system by an ensemble of
Hamiltonians whose matrix elements are independent random
variables. Dyson \cite{dyson} showed that there are three generic
ensembles of random matrices, defined in terms of the symmetry
properties of the Hamiltonian. Time-reversal-invariant quantum
system are represented by a Gaussian orthogonal ensemble ({\bf
GOE}) of random matrices when the system has rotational symmetry
and by a Gaussian symplectic ensemble ({\bf GSE}) otherwise.
Chaotic systems without time reversal invariance are represented
by the Gaussian unitary ensemble ({\bf GUE}). The dimension $\beta
$ of the underlying parameter space is used to label these three
ensembles: for {\bf GOE}, {\bf GUE} and {\bf GSE}, $\beta $ takes
the values 1, 2 and 4, respectively. Balian \cite{balian} derived
the weight functions $P_{\beta }({\bf H})$ for the three Gaussian
ensembles from the maximum entropy principle postulating the
existence of a second moment of the Hamiltonian. He applied the
conventional Shannon definition for the entropy to ensembles of
random matrices as $S=-\int $d${\bf H}P_{\beta }({\bf H})\ln
P_{\beta }(\bf {H})$ and maximized it under the constraints of
normalization and fixed mean value of Tr$({\bf H^+H}) $ and
obtained

\begin{equation}
P_{\beta }^{G}(\eta ,{\bf H})=Z_{\beta}^{-1}\exp \left[ -\eta {\rm
{Tr}\left( {\bf H^+H}\right) }\right] ,\
\end{equation}%
which is a Gaussian distribution with inverse variance $1/2\eta $. Here $%
Z_{\beta}^{-1}$ is a normalization constant.

The non-extensive {\bf RMT} applies the maximum entropy principle,
with Tsallis' entropy, to obtain matrix-element distributions that
are no more independent. The Tsallis entropy is defined for the
joint matrix-element probability density $P_{\beta }(q,{\bf H})$\
by
\begin{equation}
S_{q}\left[ P_{\beta }(q,{\bf H})\right] =\left. \left( 1-\int
{\rm d}{\bf H}\left[ P_{\beta }(q,{\bf H})\right] ^{q}\right)
\right/ (q-1).
\end{equation}%
We shall refer to the corresponding ensembles as the Tsallis
orthogonal ensemble ({\bf TsOE}), the Tsallis Unitary ensemble
({\bf TsUE}), and the Tsallis symplectic ensemble ({\bf TsSE}).
For $q\rightarrow 1$, $S_{q}\ $tends to Shannon's entropy, which
yields the canonical Gaussian orthogonal, unitary or symplectic
ensembles\ ({\bf GOE}, {\bf GUE}, {\bf GSE}) \cite{mehta,balian}.

There are more than one formulation of non-extensive statistics
which mainly differ in the definition of the averaging. Some of
them are discussed in \cite{wang}. We apply the most recent
formulation \cite{Ts3}. The probability distribution $P_{\beta
}(q,{\bf H})$\ is obtained by maximizing the
entropy under two conditions, 

\begin{eqnarray}
\int {\rm d}{\bf H} P_{\beta }(q,{\bf H})& = & 1, \label{cons1}\\
\frac{\int {\rm d}{\bf H}\left[ P_{\beta }(q,{\bf H})\right]
^{q}{\rm {Tr}\left( {\bf H^+H}\right) }}{\int {\rm d}{\bf H}\left[
P_{\beta }(q,{\bf H})\right] ^{q}}& = & \sigma _{\beta
}^{2},\label{cons2}
\end{eqnarray}
where the constant $\sigma _{\beta }$ is related to the variances
of the matrix elements. The
optimization of $S_{q}$ with these constraints yields a power-law type for $%
P_{\beta }(q,H)$, which may be expressed as \cite{abul}


\begin{equation}
P_{\beta }(q,{\bf H})=\widetilde{Z}_{q}^{-1}\left[ 1-(1-q)\widetilde{\eta }%
_{q}\left\{ {\rm {Tr}\left( {\bf H^+H}\right) -\sigma _{\beta }^{2}}%
\right\} \right] _{+}^{1/(1-q)},\label{notationplus}
\end{equation}
where $\widetilde{\eta }_{q}=\eta /\int {\rm d}{\bf H}\left[
P_{\beta }(q,{\bf H})\right]^{q}$, $\eta$ is the lagrange
multiplier associated with the constraint in Eq. (\ref{cons2}),
and $\widetilde{Z}_{q}=\int {\rm d}{\bf H}\left[ 1-(1-q)\widetilde{\eta }%
_{q}\left\{ {\rm {Tr}\left( \bf{H^+H}\right) -\sigma _{\beta }^{2}}%
\right\} \right] _{+}^{1/(1-q)}$ is a normalization constant. We
use in Eq. (\ref{notationplus}) the notation $[u]_{+}=\max
\{0,u\}$.

\bigskip

We now calculate the joint probability density for the eigenvalues
of the Hamiltonian ${\bf H}$. With ${\bf H=U^{-1}XU}$, where ${\bf
U}$\ is a unitary matrix. For this purpose, we introduce the
elements of the diagonal matrix of eigenvalues ${\bf X}=$
diag$(x_{1},\cdots ,x_{N})$ of the eigenvalues and the independent
elements of ${\bf U}$ as new variables. Then the volume element
(3) has the form
\begin{equation}
{\rm d}{\bf H}=\left\vert \Delta _{N}\left( {\bf X}\right)
\right\vert ^{\beta }{\rm d}{\bf X}d\mu ({\bf U}), \label{dh}
\end{equation}%
where $\Delta _{N}\left( {\bf X}\right) =\prod_{n>m}(x_{n}-x_{m})$
is the Vandermonde determinant and $d\mu ({\bf U})$ the invariant
Haar measure of the unitary group \cite{mehta}. In terms of the
new variables, Tr$({\bf H^+H}) =\sum_{i=1}^{N}x_{i}^{2}$ so that
the right-hand side of Eq. (\ref{dh}) is independent of the
angular variables in ${\bf U}$. \ Integrating Eq.
(\ref{notationplus}) over $\mu ({\bf U})$ yields the joint
probability density of eigenvalues in the form
\begin{equation}
P_{\beta }(q,x_{1},\cdots ,x_{N})=C_{\beta }(q)\left\vert
\prod_{n>m}(x_{n}-x_{m})\right\vert ^{\beta }\left[ 1-(1-q)\eta
_{q}\sum_{i=1}^{N}x_{i}^{2}\right] ^{\frac{1}{1-q}},
\label{P_beta_q_1}
\end{equation}%
here $C_{\beta }$ is a normalization constant.

Analytical expressions for the nearest-neighbor spacing ({\bf
NNS}) distribution can be obtained for the $2\times 2$ matrix
ensembles. For the Gaussian ensembles, this approach leads to the
well-known Wigner surmises
\begin{equation}
P_{\beta }(s)=a_{\beta }s^{\beta }\exp \left( -b_{\beta
}s^{2}\right) ,\label{Ps}
\end{equation}%
where $\left( a_{\beta },b_{\beta }\right) =\left( \frac{\pi
}{2},\frac{\pi
}{4}\right) $, $\left( \frac{32}{\pi ^{2}},\frac{4}{\pi }\right) $ and $%
\left( \frac{2^{18}}{3^{6}\pi ^{3}},\frac{64}{9\pi }\right) $ for
{\bf GOE}, {\bf GUE}, and {\bf GSE}, respectively. They present
accurate approximation to the exact results for the case of
$N\rightarrow \infty $.

We consider the non-extensive generalization of the Wigner
surmises, hoping that the results present a reasonable
approximation to the physically interesting cases of large $N$, as
they do well in the case of Gaussian ensembles. We rewrite Eq.
(\ref{P_beta_q_1}) for the case of $N=2$, and introduce the new
variable\ $s=\left\vert x_{1}-x_{2}\right\vert $ and
$X=(x_{1}+x_{2})/2$ to obtain
\begin{equation}
P_{\beta }(q,s,X)=C_{\beta }(q)s^{\beta }\left[ 1-(1-q)\eta _{q}\left( \frac{%
1}{2}s^{2}+2X^{2}\right) \right] ^{\frac{1}{1-q}}.
\label{P_beta_q_s_X}
\end{equation}%
Now, we consider the case of $q<1$. For this case, the
distribution in Eq. (\ref{Ps}) has to be complemented by the
auxiliary condition that the quantity inside the
square bracket has to be positive. Thus, we integrate over $X$ from $-z$ to $%
z$, where $z=\sqrt{1-\frac{1}{2}(1-q)\eta _{q}s^{2}}$. We obtain
the following expression for {\bf NNS} distribution in the
super-extensive regime
\begin{equation}
P_{\beta }(q,s)=a_{\beta }(q)s^{\beta }\left[ 1-b_{\beta }\left(
q\right) s^{2}\right] ^{\frac{1}{1-q}+\frac{1}{2}},
\label{P_beta_q_s}
\end{equation}%
where
\begin{equation}
a_{\beta }(q)=\frac{2\left[ b_{\beta }(q)\right] ^{\left( \beta
+1\right) /2}\Gamma \left( 2+\frac{\beta }{2}+\frac{1}{1-q}\right)
}{\Gamma \left( \frac{\beta +1}{2}\right) \Gamma \left(
\frac{3}{2}+\frac{1}{1-q}\right) },
\end{equation}%
and
\begin{equation}
b_{\beta }(q)=\left[ \frac{\Gamma \left( \frac{\beta +2}{2}\right)
\Gamma
\left( 2+\frac{\beta }{2}+\frac{1}{1-q}\right) }{\Gamma \left( \frac{\beta +1%
}{2}\right) \Gamma \left( \frac{5}{2}+\frac{\beta }{2}+\frac{1}{1-q}\right) }%
\right] ^{2}.
\end{equation}%
Thus, in the case of conserved time reversal symmetry when $\beta
=1$,


\begin{eqnarray}
a_{1}(q)&=&\left( \frac{2}{1-q}+3\right) b_{1}(q), \\
b_{1}\left( q\right) &=&\frac{\pi }{4}\left[ \frac{\Gamma \left( \frac{5}{2}+%
\frac{1}{1-q}\right) }{\Gamma \left( 3+\frac{1}{1-q}\right)
}\right] ^{2}.
\end{eqnarray}
In the absence of time reversal symmetry, $\beta =2$, and
\begin{eqnarray}
a_{2}(q) &=&\frac{4\Gamma \left( 3+\frac{1}{1-q}\right)
}{\sqrt{\pi }\Gamma \left( \frac{3}{2}+\frac{1}{1-q}\right)
}\left[ b_{2}(q)\right]
^{3/2},{\rm {\ }} \\
b_{2}\left( q\right) &=&\frac{4}{\pi }\left[ \frac{\Gamma \left( 3+\frac{1}{%
1-q}\right) }{\Gamma \left( \frac{7}{2}+\frac{1}{1-q}\right)
}\right] ^{2}.
\end{eqnarray}

\section{Numerical experiment}

\begin{figure}[tbh]
\centerline{\epsfxsize=0.9\textwidth\epsfbox{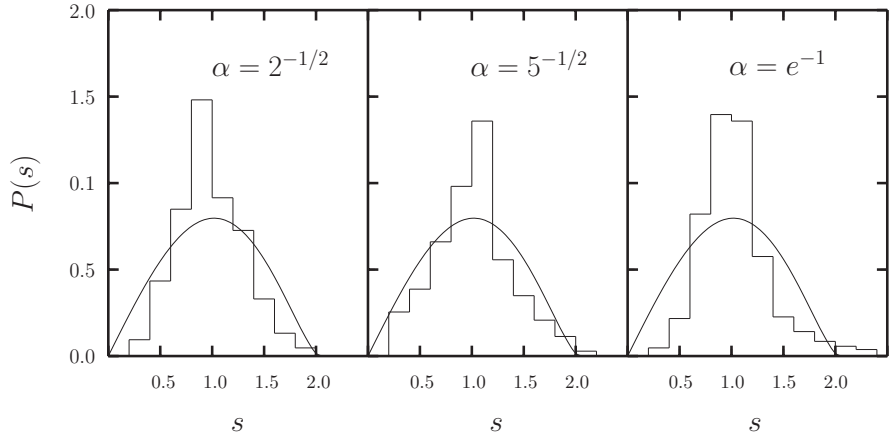}}
\caption{\protect\vspace{0cm}Comparison between {\bf NNS}
distributions for
two-dimensional harmonic oscillators with different frequency ratios $%
\protect\alpha $ obtained by Berry and Tabor \protect\cite{berry},
and the {\bf NNS} distribution of the superstatistical ensemble
with $q=0$.} \label{Fig.1}
\end{figure}

It has been suggested in \cite{abul} that the super-extensive
regime corresponds to the evolution of the chaotic system towards
a regular regime like that of a two-dimensional irrational
oscillator described by Berry and
Tabor \cite{berry}. The energy eigenvalues for such a system is given by%
\begin{equation}
E_{i}=\hbar \omega \left( n+\alpha m\right) ,  \label{E_i}
\end{equation}%
where $\omega $ is the larger frequency and $\alpha $ is the
frequency ratio while $i$ stands for the quantum-number pair $m$
and $n$. Berry and Tabor calculated 10 000 eigenvalues and
constructed the histograms of $P(s)$ for oscillators with $\alpha
=1/\sqrt{2}$, $1/\sqrt{5}$, and $e^{-1}$. These
histograms are compared in Fig. \ref{Fig.1} with the spacing distribution $%
P_{1}(0,s)$, calculated using Eq. (\ref{P_beta_q_s}). 
\newline
The figure shows that the distribution $P_{1}(0,s)$ does not
exactly fit the numerical histograms. This situation here is
similar to the case of sub-extensive regime considered in
\cite{abul,abul1}. It was found in these papers that the
non-extensive {\bf RMT} cannot reproduce the Poisson distribution
that described generic regular system for any value of $q>1$. It
can only describe the evolution of the shape spacing distribution
until its peak reaches roughly half the distance from the peak
position of the Wigner distribution to that of the Poisson.

We shall demonstrate that the super-extensive distribution
describes the final stage of a transition from a regular harmonic
dynamics to chaos in a similar way as the sub-extensive theory
describes the final stage of the stochastic transition of generic
regular systems \cite{abul,abul1}.

To show this, we shall now consider the following matrix Hamiltonian%
\begin{equation}
{\bf H}(g)=(1-g){\bf H}_{{\rm {HO}}}+g {\bf H}_{{\rm {{\bf
GOE}}}}, \label{h_g}
\end{equation}%
where ${\bf H}_{{\rm {HO}}}=$ diag$(E_{i})$ is a diagonal matrix
whose diagonal elements are given by Eq. (\ref{E_i}) while ${\bf
H}_{{\rm {{\bf GOE}}}}$\ is a random matrix of equal dimension
with entries drawn from a {\bf GOE}. By varying the parameter $g$
from 0 to 1, the Hamiltonian ${\bf H}(g)$ describes a transition
from a regular harmonic-oscillator type regime to chaotic
dynamics.

\begin{figure}[h!tb]
\centerline{\epsfxsize=0.80\textwidth\epsfbox{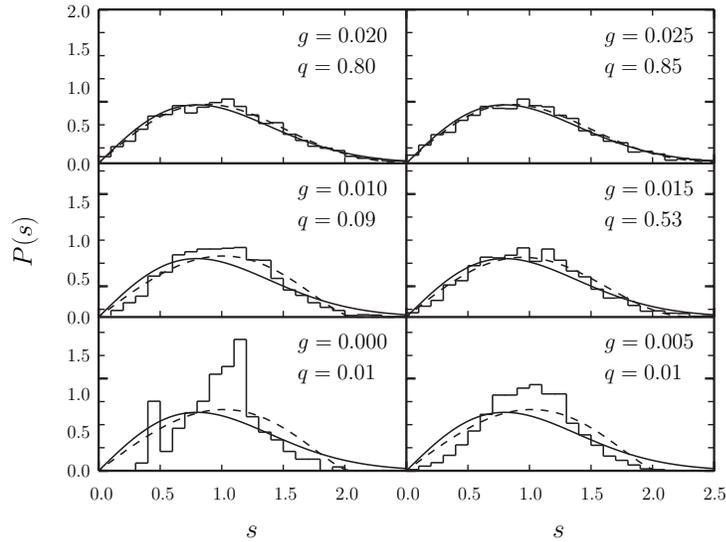}}
\caption{\protect\vspace{0.0cm}{\bf NNS} distributions for the
Hamiltonian matrix representing a two dimensional harmonic
oscillator perturbed by a {\bf GOE} Eq. (\ref{h_g}) for different
values of the parameter $g$. Dashed curves represent fitting the
histograms to Eq. (\ref{P_beta_q_s}) for $ \beta=1$. The values of
the entropic index $q$ that best fit these distributions are
given. The solid curves are pure {\bf GOE}.} \label{Fig.2}
\end{figure}

\begin{figure}[h!tb]
\centerline{\epsfxsize=0.75\textwidth\epsfbox{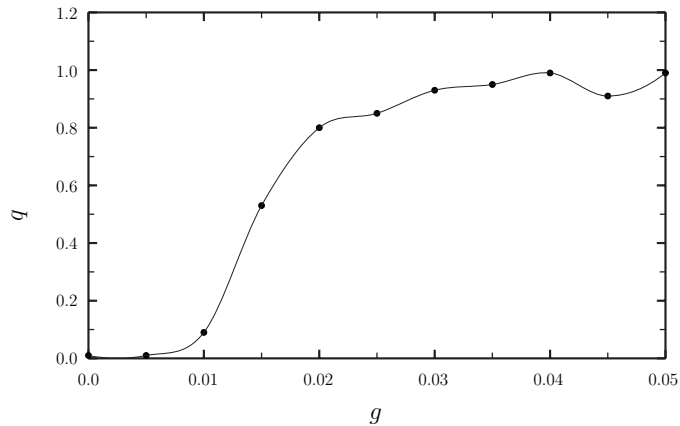}}
\caption{\protect\vspace{0.0cm} Variation the entropic index $q$
of Eq.(\ref{P_beta_q_s}) for $\beta=1$ that best fits the {\bf
NNS} distributions of the Hamiltonian in Eq. (\ref{h_g}) with the
parameter $g$.} \label{Fig.3}
\end{figure}

We have considered ensembles of 20 matrices of dimension 200
$\times$ 200 representing the Hamiltonian in Eq. (\ref{h_g}). We
numerically diagonalized these matrices and calculated the {\bf
NNS} distributions for different values of the parameter $g$. We
compared the resulting spacing distributions to Eq.
(\ref{P_beta_q_s}) for $\beta=1$ and evaluated the values of the
entropic index parameter $q$ that best fit these distributions.
The results are shown in Fig. \ref{Fig.2}. We see that for $g
\simeq 0.02$ we approach the {\bf GOE} distribution. Fig.
\ref{Fig.3} shows the variation of the values of the entropic
index $q$ that best fit these distributions with the strength of
the {\bf GOE} perturbation $g$ stating from $g=0$ where $q=0$ and
the spacing distributions take the form of picked-fence type up to
values of $g$ where $q$ approaches 1 and the spacing distributions
become that of ${\bf GOE}$.

\begin{figure}[h!tb]
\centerline{\epsfxsize=0.80\textwidth\epsfbox{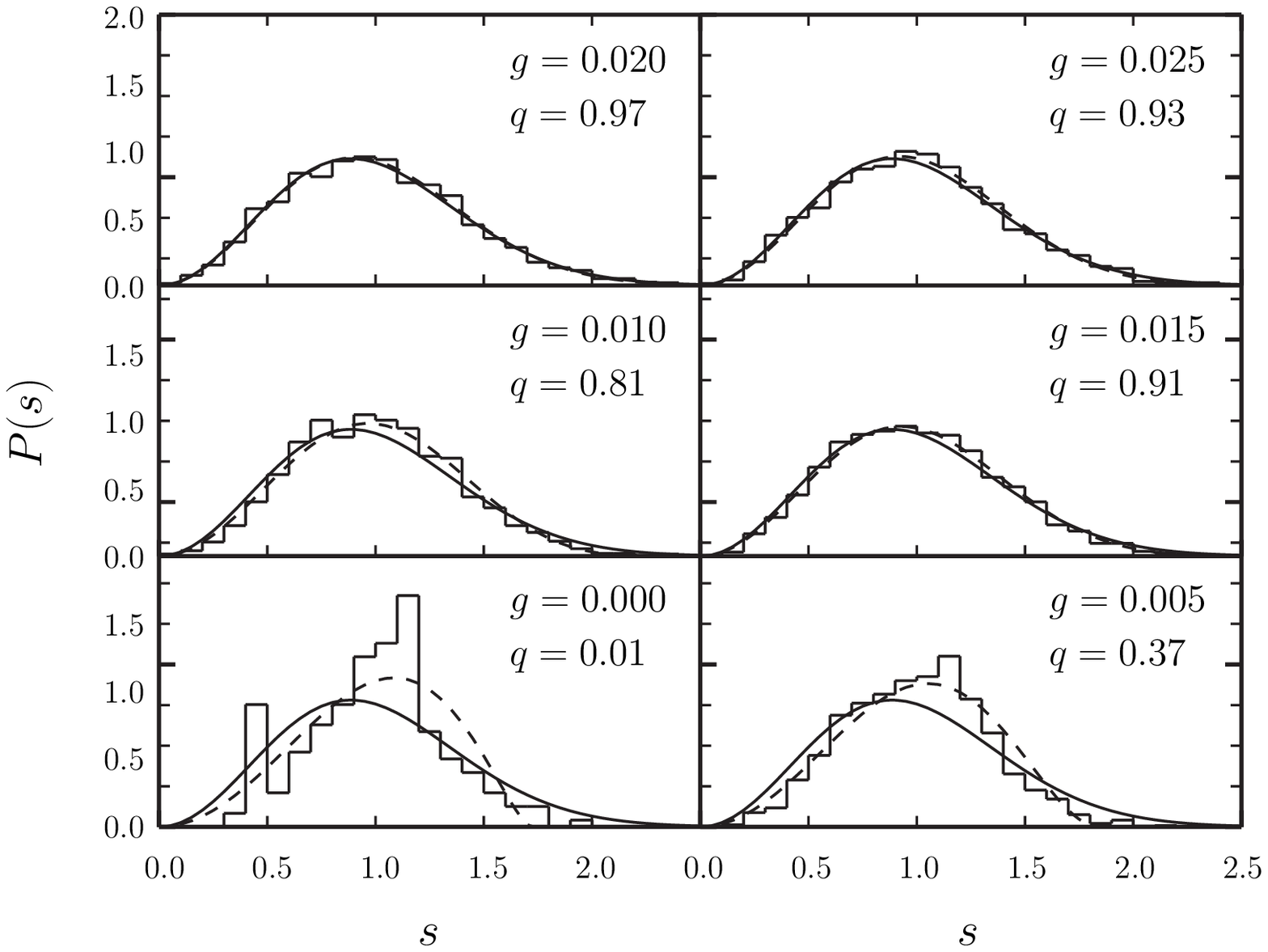}}
\caption{\vspace{0.0cm}Same as Fig. \ref{Fig.2}, but for $\beta=2$
and {\bf GUE} perturbation.} \label{Fig.4}
\end{figure}

\begin{figure}[h!tb]
\centerline{\epsfxsize=0.75\textwidth\epsfbox{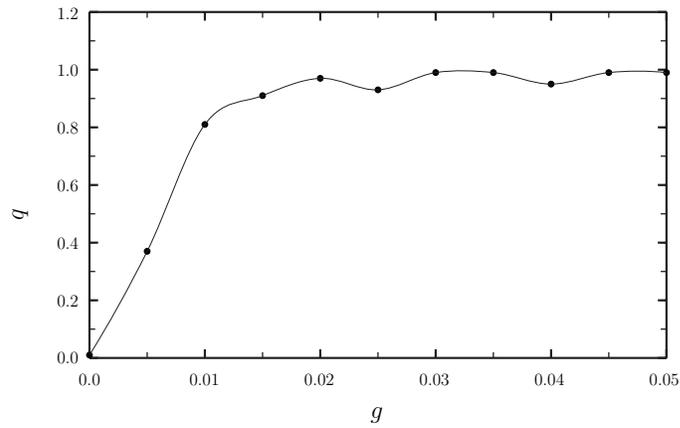}}
\caption{\vspace{0.0cm}Same as Fig. \ref{Fig.3}, but for $\beta=2$
and {\bf GUE} perturbation.} \label{Fig.5}
\end{figure}

We have considered perturbing the two-dimensional oscillator with
a {\bf GUE}, where the Hamiltonian takes the form
\begin{equation}
{\bf H}(g)=(1-g){\bf H}_{{\rm {HO}}}+g{\bf H}_{{\rm {{\bf GUE}}}},
\label{h_g_GUE}
\end{equation}%
Fig. \ref{Fig.4} shows the {\bf NNS} distributions of the
two-dimensional oscillator perturbed by a {\bf GUE} of strength
$g$. We compare the resulting spacing distributions to Eq.
(\ref{P_beta_q_s}) for $\beta=2$ and evaluated the values of the
entropic index $q$ that best fit these distributions. Fig.
\ref{Fig.5} shows the variation of the entropic index $q$ with the
{\bf GUE} perturbation strength $g$.

From Fig. \ref{Fig.2} and Fig. \ref{Fig.4}, we see that Eq.
(\ref{P_beta_q_s}) does not exactly fit the initial stage of the
transition from the regular two-dimensional oscillator dynamics
but fits the final stage of the transition into both {\bf GOE} and
{\bf GUE}. From Fig. \ref{Fig.3} and Fig. \ref{Fig.5}, we see that
the {\bf GUE} limit is approached faster than the {\bf GOE} limit.

\section{Conclusion}

In summary, the {\bf NNS} distribution obtained from applying
Tsallis statistics to {\bf RMT} maps two routes for the transition
from chaos to order. One leads
towards the integrability described by the Poisson statistics by increasing $%
q>1$, but ends at the edge of chaos. The other is followed by
decreasing $q$ form 1 to 0. It leads to a picket-fence type
spectrum, such as the one obtained by Berry and Tabor \cite{berry}
for the two-dimensional harmonic oscillator with non-commensurate
frequencies.

\newpage 


\newpage

\begin{thebibliography}{99}
\bibitem{Ts1} C. Tsallis, $\it{J.\ Stat.\ Phys.}$ 52, 479 (1988).

\bibitem{Ts2} C. Tsallis, $\it{Lect.\ Notes\ Phys.}$ 560, 3 (2001).

\bibitem{Ts23} C. Tsallis, $\it{J.\ Comput.\ Appl.\ Math.}$ 227, 51 (2009).

\bibitem{mehta} M.L. Mehta, $\it{Random\ Matrices}$, second ed., Academic Press, San Diego,
1991.


\bibitem{abul51} A.Y. Abul-Magd, $\it{Eur.\ Phys.\ J.}$ B 70, 39 (2009).

\bibitem{abul52} A.Y. Abul-Magd, B. Dietz, T. Friedrich, A. Richter, $\it{Phys.\
Rev.\ E}$ 77, 046202 (2008).


\bibitem{balian} R. Balian, $\it{Nuovo\ Cim.}$ 57, 183 (1958).

\bibitem{evans} J. Evans and F. Michael, e-print cond-mat/0207472
and e-print cond-mat/0208151.

\bibitem{toscano} F. Toscano, R.O. Vallejos, and C. Tsallis, $\it{Phys.\ Rev.\ E}$
69, 066131 (2004).

\bibitem{nobre} F.D. Nobre and A.M.C. Souza, $\it{Physica\ A}$ 339, 354 (2004).

\bibitem{bertuola} A.C. Bertuola, O. Bohigas, and M.P. Prato, $\it{Phys.\ Rev.\ E}$
70, 065102(R) (2004).

\bibitem{abul} A.Y. Abul-Magd, $\it{Phys.\ Lett.\ A}$ 333, 16 (2004).

\bibitem{abul1} A.Y. Abul-Magd, $\it{Phys.\ Rev.\ E}$ 71, 066207 (2005).

\bibitem{berry} E.V. Berry and M. Tabor, $\it{Proc.\ Roy.\ Soc.\ London,\ Ser.\ A}$
356, 375 (1977).

\bibitem{dyson} F.J. Dyson, $\it{J.\ Math.\ Phys.}$ 3, 140 (1962).

\bibitem{wang} Q.A. Wang, $\it{Eur.\ Phys.\ J.\ B}$ 26, 357 (2002).

\bibitem{Ts3} C. Tsallis, R.S. Mendes, and A.R. Plastino, $\it{Physica\ A}$
261, 534 (1988).





\end{thebibliography}
\end{document}